# Co-evolution of replicators and their parasites

Alexander Spirov

*The Institute of Scientific Information for Social Sciences RAS, Moscow, Russia*

*alexander.spirov55@gmail.com*

**ABSTRACT**

The problem of evolutionary complexification of life is considered one of the fundamental aspects in contemporary evolutionary theory. Parasitism is ubiquitous, inevitable, and arises as soon as the first replicators appear, even during the prebiotic stages of evolution. Both in theoretical approaches (computer modeling and analysis) and in real experiments (replication of biological macromolecules), parasitic processes emerge almost immediately. An effective way to avoid the elimination of the host-parasite system is through compartmentalization. In both theory and experiments, the pressure of parasitism leads to the complexification of the host-parasite system into a network of cooperative replicators and their parasites. Parasites have the ability to create niches for new replicators. The co-evolutionary arms race between defense systems and counter-defense mechanisms among parasites and hosts can progress for a considerable duration, involving multiple stages, if not indefinitely.

## Оглавление

# 1. ВВЕДЕНИЕ: КОМПЛЕКСИФИКАЦИЯ В СИСТЕМЕ ПАРАЗИТ-ХОЗЯИН

## 1.1 Проблема комплексификации в современной биологии

Вряд ли можно отрицать, что в ходе биологической эволюции сложность если не всех, то огромного большинства процессов и механизмов драматически возросла. Действительно, накопленные знания свидетельствуют о том, что на самых ранних этапах жизни организмы были организованы много проще, чем большинство современных. Так что проблема усложнения (комплексификации[1]) процессов и механизмов в эволюции остается одной из актуальных на сегодня.

Рассматривая детально пути эволюции, можно обнаружить этапы появления новых свойств, а также этапы медленного изменения существующих. Чтобы смоделировать происхождение новых функций, нам нужен механизм того, как сложное поведение более высокого уровня возникает из взаимодействий низкого уровня [Kaneko, 2006[1]]. Природа, как видно, использует малоизученные и даже еще неисследованные механизмы эволюционного усложнения.

Усложнение генетических программ построения организма из единственной исходной клетки — один из наиболее впечатляющих аспектов становления организации живых существ в эволюции. Современная биология рассматривает генетические программы в рамках концепции генетических сетей, поэтому под усложнением такой программы понимают усложнение организации генетических сетей (т.е. числа генов в ансамбле и количества функциональных связей между ними). Наша задача здесь — понять хотя бы некоторые базовые аспекты усложнения простых функциональных молекулярных ансамблей.

### 1.1.1 Ко-эволюционная гонка вооружений между паразитами и их хозяевами

Многие современные авторы предполагают, что концепция «гонки вооружений» является лучшей схемой, в которой реорганизации генетического материала и естественный отбор могут привести в течение длительного периода времени к нарастанию сложности [Dawkins, 1986[2]]. «Гонка вооружений» представляет собой усовершенствование эффективности функционирования одних агентов, позволяющее им выжить, как прямое следствие улучшения развивающихся систем функционирования других агентов. «Гонка» основана на взаимоотношениях между двумя конкурирующими организмами, «отслеживающими» изменения друг друга. Это один из примеров ко-эволюции -- взаимного влияния на эволюцию двух взаимодействующих видов.

При паразитизме взаимодействие между паразитом и его хозяином приводит к ко-эволюционной гонке вооружений, в которой эволюционный прогресс одной стороны провоцирует дальнейший ответ другой стороны. Хозяин должен разработать защитные средства, чтобы уменьшить воздействие паразитизма, в то время как паразит должен разработать средства, чтобы противостоять защите хозяина.

Математические исследования и компьютерное моделирование простых процессов репликации продемонстрировало, что в таких системах паразитные объекты и процессы неизбежны и появляются почти сразу. Более того, длительное теоретическое развитие этой проблематики показало, что ко-эволюционная гонка вооружений в таких системах способна приводить к усложнению (комплексификации) организации хозяина -- системы репликаторов. Более того, такая комплексификация наблюдалась и в реальных экспериментах (in vitro) с биохимическими репликатарами. Эти находки могут иметь критическое значение для нашего понимания биологической эволюции вообще, особенно в свете вездесущности геномных паразитов. Мы сосредоточимся здесь именно на анализе возможной роли ко-эволюционной гонки вооружений паразитов с репликаторами-хозяевами.

---

[1] Мы в этой статье будем чаще использовать термин комплексификация вместо усложнения, чтобы сделать акцент на понимании этого явления в терминах теории сложных систем и уйти от слишком широкого обыденного понимания слова усложнение.

## 2. КО-ЭВОЛЮЦИЯ ТИПА ПАРАЗИТ-ХОЗЯИН В СИСТЕМАХ С «РЕПЛИКАТОРАМИ»

Прежде чем обсуждать ключевые результаты по теме, заявленной в заглавии главы, сделаем несколько замечаний по терминологии. Под репликаторами ниже понимаются как компьютерные программы, так и каталитические ансамбли рибозимов или энзимов, способные создавать копии других таких программ или же энзимов / рибозимов, соответственно. Под «паразитами» в этой области, начиная по крайней мере с пионерских работ Эйгена (конец 70 гг прошлого века), понимают «мутантные» (т.е., измененные, неправильные) копии исходных репликаторов, которые сами не способны более реплицировать других, но которые сохраняют требуемые характеристики для того, чтобы их копировали истинные репликаторы. В простейшем случае эти сущности не паразиты в биологическом понимании, а просто «тунеядцы»: они растрачивают лимитированные ресурсы среды, не выполняя при этом функций репликатора. Далее, в этой области уделяется все большее внимание к таким событиям, когда исходная «популяция» начально идентичных репликаторов (in silico или in vitro) в ходе экспериментальной «эволюции» при посредстве «мутаций» порождает дочерние «популяции» измененных репликаторов, начальных паразитов, затем более изощренных паразитов, и так далее. На этом этапе авторы зачастую используют термины из популяционной экологии и рассуждают в базовых терминах этой науки. Надо понимать, что это отчасти научный жаргон и популяции репликаторов и «тунеядцев», как и мутуализм взаимоотношений истинных репликаторов и «тунеядцев-паразитов» не подойдут под определения популяционной биологии. Но такой жаргон упрощает описание и обсуждение результатов и заключений в этой области, если и авторы и читатели понимают эти термины сходным образом.

Проблематика процессов в системе репликаторов в связи с неизбежным появлением паразитных процессов в последние пару десятилетий исследуется не только теоретически, но и экспериментально. Она восходит к известным проблемам пребиотической эволюции вообще и Мира РНК в частности. Именно в Мире РНК мы можем ожидать появление рибозимов -- первых РНК-репликаторов. Теория и эксперимент прогнозируют почти сразу же появление паразитных молекул РНК, неспособных к репликации, но способных использовать репликаторы в своих целях. Вредоносность паразитов создает эволюционное давление на РНК-репликаторы мутировать в направлении избегания вредоносности паразитных РНК. Такая гонка вооружений может, в частности, способствовать комплексификации РНК-репликаторов с эволюционным формированием целых сетей таких репликаторов, что позволяет избегать давления паразитов. Комплексификация ансамблей РНК-репликаторов как побочный продукт эволюционной гонки вооружения с самого начала пребиотической эволюции может иметь огромное значение для происхождения молекулярных основ Жизни, какой мы ее наблюдаем теперь.

### 2.1    Ко-эволюция «репликаторов» и паразитов в моделях

В этом разделе мы используем термин репликатор обобщенно, понимая и конкретные (хотя и гипотетические) молекулы РНК, способные к копированию других РНК (включая копии себя). Здесь же мы рассмотрим и более абстрактные «репликаторы» из моделей миров (как Тиерра и Авида), населенных цифровыми организмами.

Первые существенные шаги в разработке теории ансамблей репликаторов были сделаны Эйгеном [Eigen 1989[3]]; см. также работу Эйгена и Шустера [Eigen, Schuster, 1979[4]]. Их модель "гиперцикла" предполагает, что набор самовоспроизводящихся рибозимов обеспечивают друг друга гетерокаталитической помощью в круговой топологии (т.е. репликатор A поддерживает репликацию репликатора B, B поддерживает C, C поддерживает A). Однако модель гиперцикла оказалась уязвимой перед паразитическими мутантными репликаторами: как "эгоистичные" (принимающие участие, но не предоставляющие поддержку для репликации), так и "короткозамкнутые" (поддерживающие репликацию более отдаленного члена системы) паразиты разрушают гиперциклическое сосуществование и приводят к исключению всех репликаторов, кроме одного [Smith 1979[5]]. Как было в свое время показано, практически единственный способ сохранить гиперцикл — это компартментализация: систему необходимо заключить в мембранные пузырьки, и отбор на уровне пузырьков может поддерживать сосуществование гиперциклически связанных репликаторов [Eigen et al., 1981[6]]. Более поздние исследования показали, что системы, подобные гиперциклам, способны ко-эволюционировать с паразитами в условиях пространственно распределенной системы [Könnyu et al., 2008[7]].

### 2.1.1 Паразиты в эволюции цифровых организмов

Здесь мы рассмотрим хорошо известные модели эволюции, такие как Tierra и Avida.

#### 2.1.1.1 Тиерра Томаса Рэя

Система Тьерра Томаса Рэя [Ray, 1991[8]] способствовала проведению некоторых из первых успешных экспериментов по эволюции с самовоспроизводящимися компьютерными программами. Tierra защищала «живые» программы от перезаписи конкурентами. Когда население выросло до уровня несущей способности окружающей среды, Тьерра удалила из населения самые старые программы, чтобы освободить место для рождения новых программ.

В первоначальных экспериментах с использованием Tierra Рэй создал популяции с наследственной программой, способной только к самовоспроизведению [Ray, 1991]. В этих ранних исследованиях доминировала конкуренция за пространство, что привело к сильному давлению отбора на организмы, заставившему их увеличить скорость размножения. Рэй наблюдал, как организмы с более короткими геномами, которые можно было копировать быстрее, эволюционировали и вытесняли организмы с более длинными геномами, которым требовалось больше времени для размножения.

На Тьерре Рэй неожиданно наблюдал эволюцию облигатных паразитов — программ, которые использовали копировальный аппарат своих конкурентов для копирования самих себя. Эти программы конкурировали за тот же ограниченный ресурс (пространство), что и организмы-«хозяева», косвенно нанося вред приспособленности своего хозяина, не позволяя хозяину воспроизводиться, если в окружающей среде не хватало места для потомства хозяина. В большинстве экспериментов Рэй наблюдал эволюционную гонку вооружений между хозяевами и паразитами. Потенциальные программы-хозяева развили механизмы сопротивления паразитам, а паразиты, в свою очередь, научились преодолевать эти защитные механизмы.

Впоследствии Рэй наблюдал то, что он назвал «гиперпаразитами», которые паразитировали на первоначальных косвенных паразитах. Гиперпаразит напрямую украл ресурсы (циклы ЦП) у своего хоста, обманом заставив его реплицировать геном паразита, когда хост пытался реплицировать свой собственный геном. Эти гиперпаразиты в конечном итоге эволюционировали и стали облигатно сотрудничать с другими гиперпаразитами, полагаясь на копировальный аппарат своих соседей для успешного размножения. Наконец, появились мошенники, которые не имели копировальной техники, а вместо этого пользовались преимуществами копирующей техники соседей и в результате размножались быстрее.

Таким образом, Тиерра была первой моделью компьютерной жизни, где удалось наблюдать комплексификацию ансамблей само-репродуцирующихся цифровых организмов. Компьютерный эксперимент, начинавшийся с популяции копий единственной способной к самокопированию программы (сделанной «вручную»), приводил в итоге к появлению паразитических и гипер-паразитических программ, связанных друг с другом и в итоге зависящих от функционирования программ-хозяев.

Богатство наблюдаемой эволюционной динамики на Тьерре поначалу было удивительным, учитывая простоту окружающей среды Тьерры. Однако исследования цифровой эволюции на тот момент носили в основном наблюдательный характер. Следующим этапом в исследованиях цифровой эволюции станет интеллектуальный преемник Тьерры.

### 2.1.1.2    Ко-эволюция в модели Авида

Платформа Avida Digital Evolution расширила дизайн Tierra, добавив перекрестное поведение и настоящий паразитизм, а также набор сложных инструментов отслеживания данных и возможность исследователям настраивать сложные среды (Adami et al, 2000[9]; Ofria et al., 2009[10]). Благодаря этой поддержке цифровая эволюция вышла за рамки наблюдательных исследований естественной истории виртуальных организмов и перешла к контролируемым экспериментальным исследованиям.

В Avida самовоспроизводящиеся «цифровые организмы» конкурируют за место в решетке пространств сетки [Ofria et al., 2009]. Когда организм размножается, его потомство помещается в соседнее пространство (или в случайное пространство, если популяция хорошо смешана), заменяя всех обитателей этого пространства. Как и в Tierra, повышение скорости самовоспроизведения выгодно во время конкуренции за место в окружающей среде, и организмы в Avida могут увеличивать скорость репликации за счет повышения эффективности репликации (например, используя более компактное кодирование). Кроме того, Авида представила концепцию ресурсов, которые могут «метаболизироваться» цифровым организмом, чтобы ускорить скорость, с которой он выражает свой геном (т. е. его «скорость метаболизма»). Ресурсы в Avida связаны с выполнением определенных задач, таких как вычисление функций логической логики на входных данных из среды.

Авида реализует концепцию ресурсов, которые могут «метаболизироваться» цифровым организмом, чтобы ускорить скорость, с которой он экспрессирует свой геном. Ресурсы в Avida связаны с выполнением определенных задач, таких как вычисление Булевых логических функций на входных данных из среды.

Симбиоз стал возможен в Avida, когда Zaman et al. [2011] добавили поддержку паразитов. Эти паразиты являются облигатными эндосимбионтами, которые крадут заданное пользователем количество ресурсов (в виде циклов ЦП) у своих хозяев. Паразиты могут горизонтально передавать свое потомство другим хозяевам, копируя свой геном, а затем пытаясь внедрить свое потомство другому хозяину. Однако инъекция успешна только в том случае, если паразит выполняет одну из тех же задач, что и потенциальный хозяин.

Используя платформу Avida для цифровой эволюции, Zaman с соавторами [Zaman et al., 2014[11]] показывают, что ко-эволюция хозяев и паразитов значительно увеличивает сложность (комплексность) организмов по сравнению с той, которая достигается в противном случае (без паразитов). В этих численных тестах популяции самореплицирующихся компьютерных программ-хозяев ко-эволюционируют вместе с паразитными программами, которые отбирают вычислительную мощность у своих хозяев. По мере того, как паразиты эволюционируют так, чтобы преодолевать резистентность хозяев, хозяева, как следствие, развивают более сложные функции чтобы противостоять давлению паразита. Впечатляет то наблюдение, что хозяева вследствие ко-эволюции с паразитами приобретают геномы, которые также фенотипически более подвержены эволюционным изменениям, то есть их эволюционируемость увеличивается. Более того, авторы обнаружили, что для того, чтобы совместная эволюция способствовала развитию комплексификации, должно возникать множество совместно эволюционирующих линий.

Эволюционная комплексификация хозяев выражается в усложнении тех Булевых логических функций на входных данных из среды, необходимых для метаболизма. Ко-эволюция с паразитами (не до конца ясным образом) способствует усложнению этих логических функций.

Поскольку ко-эволюция является распространенным явлением в природе, такие результаты подтверждают общую модель, согласно которой антагонистические взаимодействия и естественный отбор вместе способствуют как увеличению сложности, так и способности к эволюции. Одной из гипотез является та, что антагонистическая ко-эволюция между хозяевами и паразитами может стимулировать эволюцию более сложных черт путем содействия гонкам вооружений с увеличением систем защиты и контрзащиты.

### 2.1.1.3 Ничто в эволюции не имеет смысла, кроме как в свете паразитизма

Недавняя публикация команды Hogeweg [Hickinbotham et al., 2021[12]] как бы заключает на сегодняшний день ключевые публикации по теме близкой к цифровым организмам и репликаторам одновременно. В этой работе авторы исследуют эволюцию репликаторов средствами моделей / платформ / программного обеспечения автоматной химии (пакет Stringmol). Каждый репликатор представляет собой короткую компьютерную программу, состоящую из последовательности операций, которая может реплицировать другие строки (программы). Эти строки-репликаторы не копируют сами себя, но копируют другие строки, которые могут быть копиями этих репликаторов, но также могут быть и нерепликаторами. Таким образом, эти строки напоминают РНК-репликазы гипотетического мира РНК в пребиотической эволюции [Joyce, Orgel, 1999[13]]. В таких системах появление "паразитов", которые реплицируются, но не обладают способностью реплицировать себя или других, считается неизбежным.

В отличие от большинства моделей мира РНК (но, как и сама реальная репликация РНК), воспроизведение в Stringmol является активным процессом, требующим времени при выполнении программы и копирующим коды элемент за элементом. Это должно сильно не способствовать эволюции в направлении более длинных, более сложных репликаторов и наоборот сильно способствовать эволюции быстрореплицирующихся "паразитов", которые, потеряв код репликации, склонны быть (гораздо) более короткими.

Ранние эксперименты с Stringmol быстро приводили к вымиранию всех строк из-за конкуренции с быстрореплицирующимися паразитами. Такая эволюция в сторону исчезновения системы паразит-хозяин вполне ожидаема. Этого вымирания можно избежать (как известно) если репликаторы распределены по ячейкам или в пространстве. Тогда формирование пространственных паттернов репликаторов и паразитов и, как следствие, более сложный отбор предотвращает вымирание. В этой публикации авторы встраивают Stringmol в «пространство» и изучают, как механизм явной репликации (средствами Stringmol) эволюционирует, чтобы справиться с появлением быстрореплицирующихся паразитов.

Система инициализируется созданным вручную репликатором, который копирует другие репликаторы с небольшой вероятностью точечной мутации. Практически сразу возникают более короткие паразиты; они копируются быстрее и поэтому имеют эволюционное преимущество. На ранних этапах эволюции, вымирание все равно происходит весьма часто из-за практически мгновенно появляющихся быстрореплицирующихся паразитов. Однако в случае, если система выживает достаточно долго для того, чтобы сформировалась хорошо известная пространственная волнообразная структура репликаторов и паразитов {replicator–parasite wave pattern}, начинается увлекательная долгосрочная эволюция.

Сложные механизмы репликации эволюционируют, чтобы справиться с паразитами. Репликаторы становятся короче и поэтому реплицируются быстрее; они развивают механизм замедления репликации, который уменьшает разницу в скорости репликации репликаторов и паразитов. Они также вырабатывают явные механизмы для различения копий себя от паразитов.

Развиваются разнообразные «экосистемы» с относительно длинными репликаторами, умеренными паразитами и большой «плотностью популяций». Репликаторы могут даже подавить паразитов настолько, что паразиты становятся редкими, а затем репликаторы, не нуждаясь больше в защите от паразитов, теряют свои сложные противодействия и становятся более простыми и короткими; но появляются новые паразиты, обратно изменяя эту тенденцию. Таким образом, мы видим, что паразиты, традиционно рассматриваемые как угроза эволюции репликаторных систем [Smith, 1979], на самом деле являются средством, с помощью которого может развиваться сложность, при условии, что формирование пространственных паттернов предотвращает глобальное вымирание.

Новые виды паразитов постоянно возникают из мутировавших репликаторов, а не из развивающихся линий паразитов. Эволюция идет эффективно за счет увеличения частоты точечных мутаций и создания новых эмерджентных мутационных операторов. Таким образом, паразитизм управляет эволюцией сложных репликаторов и сложных экосистем.

### 2.1.2 Комплексификация в системах РНК-репликаторов

#### 2.1.2.1 Стабильные ко-эволюционные режимы для генетических паразитов и их хозяев

Мы начнем наш обзор с публикации Березовской с соавторами [Berezovskaya et al., 2018[14]], где они аналитически нашли условия стабильных ко-эволюционных режимов для генетических паразитов и их хозяев-репликаторов. Методами бифуркационного анализа авторы исследовали устойчивость простых моделей ко-эволюции репликаторов и их паразитов в хорошо перемешанной среде. Они показали, что стабильная ко-эволюция хозяина и паразита (нетривиальное равновесие) возможна в версии модели, где паразит качественно отличается от хозяина-репликатора тем, что репликация паразита зависит исключительно от доступности хозяина, а не от емкости (carrying capacity) среды.

С биологической точки зрения успешный паразит должен полагаться на хозяина не только для репликации, но также для получения строительных блоков и энергии. Возможно, эти результаты в некоторой степени объясняют, почему, никакие генетические паразиты никогда не приобретали ни собственные полноценные пути биосинтеза, включая систему трансляции, ни собственные молекулярные механизмы производства энергии [Berezovskaya et al., 2018].

#### 2.1.2.2 Эволюция пребиотических репликаз в метаболической системе, связанной с минеральной поверхностью

Könnyu с соавторами [Könnyu et al., 2008] попытались представить правдоподобный сценарий первых этапов пребиотической эволюции, исследуя (средствами компьютерного моделирования) системы репликаторов, связанных с минеральной поверхностью, что должно бы благоприятствовать примитивному метаболизму (в частности, предполагается, что минеральная поверхность способна катализировать репликацию РНК). Их подход отличается от модели гиперцикла тем, что предполагается наличие рибозимных РНК-репликаторов, способных к тому же быть катализаторами в сети химических реакций, которая, в свою очередь, производит мономеры для репликации самих рибозимов (то есть такие рибозимы способны функционировать и как репликаторы и катализировать некоторые метаболические процессы). Таким образом рибозимы вносят свой вклад в производство общего ресурса (через метаболическую сеть) благодаря своей специфической каталитической активности, и метаболизм в свою очередь способствует репликации рибозимов, предоставляя мономеры.

Модель реализована как пространственно распределенная – состоящая из ячеек на плоскости, так что каждая ячейка может быть занята репликатором и возможна «диффузия» между соседними ячейками. Основная метаболическая машина - это клеточный автомат и он состоит из квадратной сетки размером 300 × 300 ячеек с тороидальной топологией, чтобы избежать краевых эффектов. Каждая ячейка сетки может содержать не более одной молекулы репликатора, которая может быть либо одним из метаболических "ферментов" («сотрудничающих» репликаторов), либо паразитом (репликатором без метаболической функции).

Авторы продемонстрировали, что эта система жизнеспособна (т.е. паразиты могут сосуществовать с хозяевами) без компартментализации мембранами, так что она поддерживает низкое количество паразитических мутантов. Более того, численные эксперименты показали, что сосуществующие паразиты системы могут служить промежуточными формами для последующей эволюции в новые рибозимы, потенциально приносящие значительную пользу метаболической системе в целом. Конкретно в данном исследовании авторы наблюдали случай эволюции паразитов в направлении функции репликатазы.

Паразитические РНК в этой системе постоянно мутируют. Более вредоносные паразиты быстро вымирают вместе с хозяйской системой. Большинство мутантов будут около-нейтральными, то есть такими же паразитическими, как и их предки, не причиняя системе ни большего, ни меньшего вреда, чем просто используя метаболизм для получения мономеров и используя их для собственной репликации. Далее такие мутанты разнообразятся и, таким образом, "сканируют" пространство последовательностей, и они все будут сосуществовать с системой-хозяином так же, как и их предки. Это означает, что множество нейтральных мутантов различных последовательностей накапливаются внутри метаболической системы по мере появления новых мутаций. Наконец, некоторые из множества нейтральных мутантов могут мутировать во что-то, что потенциально может быть полезным для самой метаболической системы – и это "что-то" может быть множеством разных по функциональности молекул РНК.

Конкретно авторы исследовали случай, когда некоторые мутанты могут проявлять немного более высокую активность репликации, чем «базовый» катализ репликации (обеспечиваемый минеральной поверхностью, на которой функционирует метаболическая система). Очевидно, что такая новонайденная в эволюции репликаза обеспечивает положительный отбор на уровне всей метаболической системы, потому что она увеличивает скорость репликации всех репликаторов, присутствующих в системе. При этом, косвенно это выгодно самой репликазе за счет селективного преимущества всей этой системы паразит-хозяин.

В целом, модель исследует возможность эволюции репликаторов и метаболизма на ранних стадиях эволюции жизни и показывает, как паразиты могли служить источником адаптивной эволюции в этой системе.

При этом следует отметить в заключение, что авторы не приводят примеров эволюционного происхождения в численных экспериментах репликазы из паразитных молекул. Результаты авторов лишь демонстрируют такую принципиальную возможность в рамках такого подхода.

### 2.1.2.3 Эволюция комплексности систем РНК-репликаторов

Группа Hogeweg [Takeuchi, Hogeweg, 2008[15]] предприняла несколько эффективных попыток исследовать проблему эволюционной комплексификации систем репликаторов подходами экспериментов in silico. Эта их модель, как и предыдущая от Könnyu с соавторами, является пространственно распределенной. В модели в явном (хотя и упрощенном) виде имплементирован ряд базовых свойств функциональных молекул РНК. Простота этой модели позволяет в явном виде моделировать отображение генотип-фенотип на уровне индивидуальных репликаторов (сиквенсов РНК). В частности, модель предполагает, что взаимодействия между репликаторами (для репликации мишени или для того, чтобы быть реплицированными) зависят от их вторичных структур и комплемнентарности их сиквенсов.

А именно, отображение генотип-фенотип в модели трактуется как сворачивание (фолдинг) РНК, где характеристики и особенности первичной структуры РНК (генотип) определяют ее вторичную структуру (фенотип). Будет ли данная молекула из взаимодействующей пары репликатором для второй молекулы или наоборот это зависит от их вторичных структур и комплементарности последовательностей между их «висячими концами» (dangling-end).

Модель Takeuchi и Hogeweg является пространственно распределенной (2D) агентной моделью на основе методов Монте-Карло. Она включает последовательности РНК длиной 50, размещенные на квадратной решетке (размером 512 × 512) с тороидальными границами, где каждый квадрат (ячейка) может содержать не более одной молекулы. Вторичная структура каждой последовательности РНК (фолдинг РНК) оценивается в модели пакетом Vienna RNA. Динамика модели — это последовательные шаги реакции и диффузии (молекулы РНК диффундируют и локально взаимодействуют). В модели формирование комплексов происходит через связывание 5'- и 3'-висячих концов двух молекул, расположенных в решетке рядом. Прочность связывания определяется уровнем комплементарности между двумя висячими концами.

Реакция репликации может происходить в комплексе пары молекул при определенных условиях: молекула, связанная своим 5'-висячим концом со второй молекулой, может ее реплицировать, если она характеризуется пре-детерминированной вторичной структурой катализатора и, если комплекс находится рядом с пустой ячейкой. Вторичная структура, способная к катализу, задана в достаточно общем виде как комбинация трех типичных элементов вторичной структуры (loops).

Шаг диффузии реализован как вариант случайного блуждания: каждая молекула может случайным образом перемещаться в одну из восьми соседних ячеек, если эта ячейка пуста. В итоге молекулы диффундируют и взаимодействуют локально.

Результаты исследования показали, что популяция репликаторов, изначально состоящая из одного генотипа, эволюционирует, образуя сложную экосистему из четырех видов-репликаторов. В ходе такой диверсификации виды эволюционируют, приобретая уникальные генотипы с различной «экологической» функциональностью. Анализ этой диверсификации показывает, что паразитические РНК способствуют эволюционному разнообразию, создавая новые "ниши" для репликаторов. Это также делает такую систему репликаторов малочувствительной к дальнейшей эволюции паразитов. Результаты также показывают, что стабильность системы в значительной степени зависит от характеристик пространственного паттерна систем репликаторов.

Отметим в заключение следующее. Хотя подход к моделированию у Takeuchi и Hogeweg во многом сходен с таковым Kamiura с соавторами (предыдущий подраздел), эта модель более детальна и ее результаты выглядят более убедительными и более близкими к реальности.

### 2.1.2.4 Эволюционная комплексификация сети молекулярной репликации с паразитами

Публикация Камиуры с соавторами [Kamiura et al., 2022[16]] во многом уникальна. Эти авторы прилагают и адаптируют модель Takeuchi и Hogeweg так, чтобы она исходила из реальных биохимических экспериментов этой команды с реальными системами репликаторов. Так что цель работы – глубже исследовать реальные экспериментальные системы репликации методами in silico.

В предшествующих экспериментальных исследованиях эта команда показала, что РНК-хозяева (РНК-репликаторы) и паразитические РНК диверсифицировались в несколько линий путем эволюции in vitro [Furubayashi et al., 2020]. В недавнем исследовании они продолжили последовательные процессы репликации и обнаружили, что диверсифицированные виды РНК начинают совместно реплицироваться, образуя взаимозависимую сеть, которая в конечном итоге состоит из трех хозяев и двух паразитов ([Mizuuchi et al., 2022], см подраздел ниже). Эти экспериментальные результаты подтверждают идею о том, что ко-эволюция между репликаторами-хозяевами и паразитическими РНК может способствовать диверсификации и комплексификации системы. Авторы ставили своей целью прояснить, каким путем такая комплексификация достигается и как обходится принцип конкурентного исключения среди видов РНК. Согласно предыдущему теоретическому исследованию Такеучи и Хогевег ([Takeuchi, Hogeweg, 2008], подраздел выше), такая сложность объясняется последовательной эволюцией паразитического вида, который создает нишу для нового вида-хозяина. В этом исследовании Камиура с соавторами тестами in silico анализировали возможные пути и механизмы процессов комплексификации, исходя из их предыдущих экспериментальных результатов.

Для этого авторы сначала построили теоретическую модель системы репликации хозяев и паразитов, которая концептуально аналогична разработанной Такеучи и Хогевег, но адаптирована к формату их конкретной экспериментальной системы. Затем они исследовали параметрическое пространство такой компьютерной модели, с целью найти режимы устойчивой репликации все видов (и паразитов и хозяев) в сетях, содержащих до трех участников. Они также провели эволюционное моделирование, введя новые репликаторы с разными параметрами. Эти симуляции согласовывались с результатами предыдущего исследования Такеучи и Хогевег, показывающего, что наиболее вероятный путь комплексификации начиная с единственного хозяина-репликатора включает последовательное добавление сначала паразита, а затем нового хозяина, устойчивого к паразиту. В итоге авторы пришли к заключению, что эти результаты моделирования объясняют детали их предыдущих реальных экспериментов.

### 2.1.3 Как Тьюринговские паразиты расширяют ландшафт цифровой жизни

С появлением жизни на Земле обработка информации приобрела несравненную важность [Szathmary, Maynard-Smith, 1997[17]; Joyce, 2002[18]]. Быстро возникли механизмы исправления ошибок, памяти (и, следовательно, зависимости от пути и случайности) [Jablonka, Lamb, 2006[19]] и способности предсказывать окружающую среду. Эти процессы тесно связаны с появлением автономных агентов [Woese, 2002[20]].

Гипотеза о том, что паразиты оказывают давление, способствующее появлению более сложных хозяев, обсуждалась выше с разных сторон. Seoane и Sole [Seoane, Sole, 2023[21]] искали самое простое и абстрактное математическое описание, которое могло бы реализовать в численных экспериментах эту качественную гипотезу.

Эти авторы использовали формальное описание паразитов и их хозяев: абстрактные машины считывают двоичную строку, реагируя на входные биты с помощью простого внутреннего алгоритма [Seoane, Sole, 2018[22]]. Такая модель вдохновлена моделью вычислительной машины Тьюринга, где абстрактное вычислительное устройство используется для формализации вычислений в терминах простого автомата, считывающего двоичную ленту.

Авторы опирались на модель "угадывания битов" [Seoane, Sole, 2018], чтобы включить в нее паразитов и рассмотреть общие вопросы ко-эволюции. "Угадыватели битов" представляют собой минимальную модель, в которой сложность (согласно информационной теории и информатике) связана с дарвиновским отбором организмов, которые процветают и реплицируются, если успешно предсказывают свою среду.

В отличие от этого исходного формализма, такие Тьюринговские модели работают в стохастических средах, которые им необходимо предсказывать. При определенных условиях (которые требуют достаточно успешной способности предсказания) они могут воспроизводить сами себя. Более того, они могут сокращать или расширять свои вычислительные возможности, если допускается их эволюционное изменение. Как показали авторы, даже такое минималистическое представление этих виртуальных агентов демонстрирует, что паразиты могут увеличивать сложность (комплексифицировать) как самих хозяев, так и себя. Предлагаемая в статье формализация среды, хозяев и паразитов как строк из битов с минимальными правилами приближает такую модель к областям информатики, информационной теории и статистической физики.

Развиваемая минималистическая модель показывает, что: i) увеличение поведенческой сложности может быть действенной стратегией для ухода от паразитов; ii) более сложные организмы легко возникают при введении простых, неизменных паразитов в экосистему; iii) эко-эволюционная динамика может привести к динамике Красной Королевы, при которой хозяева и паразиты становятся более сложными, чтобы уходить друг от друга. Эти результаты дополняют другие явные факторы, способствующие увеличению биологической сложности [Seoane, Sole, 2018], а также они ослабляют более нейтральные точки зрения, которые предполагают, что нет явного эволюционного давления, способствующего увеличению сложности в биологии. Разумно предположить, что эти силы (в частности, паразитизм) действовали с самых ранних периодов истории жизни. Это предоставляет убедительные аргументы в пользу того, что сложная жизнь ожидается в различных условиях.

Результаты, связанные с динамикой Красной Королевы, являются самыми важными выводами. Широкий набор условий приводит к быстрому, взлету сложности. Ограничивающие аспекты этих благоприятных условий (например, то, что хозяева не должны быть загнаны в вымирание и что у паразитов должны быть значительно более низкие метаболические и репликационные издержки) согласуются с реальными отношениями паразит-хозяин.

## 2.2　Эволюция репликаторов в реальных экспериментах

Безусловно наиболее впечатляющим в этой современной области молекулярной эволюции является то, что давно известные ожидания и опасения в отношении роли вездесущих паразитов в системах репликаторов удалось наблюдать в реальных биохимических экспериментах с реальными репликаторами. Более того, в подобных экспериментах удалось пронаблюдать эволюционную комплексификацию в направлении от единственного вида молекулы-катализатора к целому каталитическому ансамблей.

### 2.2.1.1 Эволюционная комплексификация каталитических ансамблей рибозимов самосплайсинга

К одним из самых ранних наблюдений комплексификации ансамблей рибозимов в экспериментальной молекулярной эволюции следует отнести работу Hanczyc и Dorit [Hanczyc, Dorit, 1998[23]] с рибозимами группы I Tetrahymena thermophila (рибозимы самосплайсинга). В процессе экспериментальной эволюции вариантов этого рибозима с целью улучшения его функций (DNA cleavage) эти авторы столкнулись с таким наблюдением. В эксперименте неожиданно появились молекулы-производные от исходного вида рибозима, способные действовать в качестве партнера для рибозима группы I, но уже не способные к автокаталитической активности. Этот новый вид РНК имеет делецию в каталитическом ядре и участвует в продуктивном межмолекулярном взаимодействии с активным рибозимом, обеспечивая тем самым свое выживание в популяции. Таким образом, в этих экспериментах по эволюции рибозимов исходная однородная популяция функционально автономных молекул породила несколько новых молекулярных видов, участвующих в новых и более сложных межмолекулярных взаимодействиях [Hanczyc, Dorit, 1998]. Отметим, что в этих экспериментах наблюдалась комплексификация от единственного вида рибозима до их ансамбля, но это были не репликаторы.

### 2.2.1.2 Автоколебания системы РНК-репликатор и паразитная РНК

На сегодняшний день системы саморепликации с использованием ДНК, РНК и белков успешно воспроизводятся in vitro [Lincoln, Joyce, 2009[24]; Cheng, Unrau, 2010[25]; Vaidya, et al., 2012[26]]. Следующие важные задачи включают воссоздание интерактивных сетей самовоспроизводящихся видов молекул и изучение того, как такие взаимодействия порождают сложное «экологическое» поведение.

Работа Bansho с соавторами [Bansho et al., 2016[27]] по сути является воспроизведением в реальном биохимическом эксперименте взаимодействий РНК-репликаторов и паразитных РНК. Авторы воссоздали in vitro простую молекулярную "экосистему", включающую два вида молекул РНК. Первая РНК – это вид РНК, производящий через внеклеточную систему трансляции РНК-зависимую РНК-полимеразу (Qβ replicase) для репликации этой РНК. Второй вид — это паразитическая РНК, которая также зависит от этого же фермента для своей репликации.

Авторы обнаружили, что паразитарная РНК уничтожает РНК-хозяина. Однако, когда система компартметализована, возникает непрерывный колебательный режим в популяционной динамике этих двух РНК. Компартментализация достигалась в эксперименте тем, что эти молекулярные ансамбли помещали в микро-капли и тем самым изолировали. При этом, характер осцилляций менялся по мере эволюции РНК-хозяина. Эти результаты демонстрируют, что компартментализация играет важную роль в ко-эволюции молекул-хозяев и молекул-паразитов, и предполагают, что происхождение ко-эволюции хозяина и паразита может восходить к самым ранним стадиям эволюции жизни [Bansho et al., 2016].

### 2.2.1.3 Возникновение и диверсификация экосистемы РНК-хозяев паразитных РНК

В предыдущем исследовании (предыдущий подраздел, [Bansho et al., 2016]) эта исследовательская команда провела эксперимент по серийному переноса вышеупомянутой системы репликации РНК для изучения ко-эволюционного процесса репликатора-хозяина и паразитических РНК. Авторы наблюдали, что РНК хозяина приобрела определенный уровень устойчивости к паразитам в заключительных раундах эксперимента по репликации (43 раунда, 215 часов). Однако они не наблюдали контр-адаптивной эволюции паразитической РНК к РНК хозяина, так что ко-эволюционный процесс хозяйских и паразитических РНК оставался не проясненным.

В следующем исследовании [Furubayashi et al., 2020[28]] эксперимент по репликации был продлен еще на 77 раундов (43+77=120 раундов, 385 часов), в надежде что более длительная ко-эволюция приведет к еще более нетривиальным последствиям.

Действительно, в ходе эксперимента в 120 раундов исходная система репликатора и его паразита превратилась в эволюционирующую экосистему паразитных и хозяйских видов РНК за счет постоянного появления новых типов РНК, возникающих в результате ошибок репликации. Хозяйские и паразитные РНК диверсифицировались, по крайней мере, в две и три разные линии соответственно, и они демонстрировали ко-эволюционную динамику гонки вооружений. Паразитная РНК накапливала уникальные мутации, тем самым добавляя новую генетическую вариацию ко всему ансамблю репликаторов. Эти результаты предоставляют впечатляющее экспериментальное свидетельство того, что ко-эволюционное взаимодействие между молекулами хозяина и его паразита играет ключевую роль в создании разнообразия и сложности в пребиотической молекулярной эволюции [Furubayashi et al., 2020].

### 2.2.1.4 Эволюционный переход от одного РНК-репликатора к сети репликаторов (Ichihashi 2022)

Долгосрочный эксперимент (120 раундов эволюции in vitro, 600 ч), описанный в предыдущем подразделе, продемонстрировал последовательное появление новых линий РНК-хозяина и паразитических РНК, демонстрирующих защитные и контрзащитные свойства [Bansho et al., 2016; Furubayashi et al., 2020]. Однако эти линии поочередно и лишь временно доминировали в популяции, возможно, из-за все еще недостаточной длительности эксперимента.

В следующей публикации [Mizuuchi et al., 2022[29]] эксперимент по серийной переносу был продолжен до 240 раундов (1200 ч). В ходе эксперимента РНК диверсифицируется в несколько сосуществующих линий хозяев и паразитов, частоты которых в популяции первоначально колеблются и постепенно стабилизируются. Последняя популяция, состоящая из пяти линий РНК, образует сеть репликаторов с разнообразными взаимодействиями, включая кооперацию, помогающую репликации всех остальных видов. Анализ сиквенсов показал, что две ранее обнаруженные линии РНК хозяина стали устойчивыми и в дальнейшем разошлись на множество сублиний РНК хозяина и паразитических РНК. Популяционная динамика каждой линии постепенно менялась в ходе эволюции: от динамически изменяющихся стадий к квазистабильному сосуществованию, что позволяет предположить появление ко-репликативных отношений между линиями. Биохимический анализ подтвердил совместную репликацию доминантных РНК в различных линиях, содержащих кооперативную РНК, которая реплицирует всех остальные виды, создавая таким образом сеть множественных репликаторов.

Эти результаты подтверждают способность молекулярных репликаторов спонтанно усложняться в ходе ко-эволюции, что является критическим шагом для возникновения жизни [Mizuuchi et al., 2022]

Как обсуждалось в предыдущем разделе, результаты этой экспериментальной работы вдохновили авторскую команду на разработку и исследование компьютерной модели эволюции репликаторов с паразитами (см предыдущий раздел).

## 3. ЗАКЛЮЧЕНИЕ

### 3.1 Паразитизм вездесущ и неизбежен

Как обсуждалось выше паразитизм в живой природе вездесущ. Более того, как только мы тем или иным способом реализуем копирование или само-копирование сиквенсов, длина и конкретная последовательность которых имеет некий смысл в нашей реализации, так сразу мы сталкиваемся с появлением паразитных сиквенсов, использующих чужие ресурсы и способности к копированию.

Это наблюдается и в простых компьютерных моделях, и в простых экспериментальных системах in vitro. Такой паразитизм выглядит почти изначальным и наблюдается уже в достаточно простых системах, весьма далеких от живого.

Вместе с тем, в реальных экспериментах получены простейшие химические системы с элементами автокатализа. Исследование их поведения пока что не доказало их способность к дарвиновской эволюции (в достаточно полной мере), но и паразиты в таких системах в явном виде пока что не описаны. Мы можем (пусть предварительно) заключить, что слишком простые авто-каталитические системы еще не способны продемонстрировать ни эффективной эволюции ни появления паразитов. И то и другое, как полагают, требует выраженных процессов переноса информации (а не просто самовоспроизведения).

### 3.2    Ко-эволюция с паразитами за пределами биологии

Эти результаты подразумевают, что мощные силы, лежащие в основе общих биологических взаимодействий, будут двигать жизнь к большой сложности при соответствующих условиях. По нашему мнению, этот устойчивый механизм переворачивает исходный вопрос ("существуют ли давления на сложную жизнь в дарвинизме") с ног на голову. Теперь нам следует задаться вопросом, какие могут быть последствия такой сильной эволюционной динамики в реальном мире или в каких условиях этот механизм может быть ослаблен или использован. В этом отношении мы не говорим ничего о явной реализации, с помощью которой может быть достигнута эта сложность (например, составные фенотипы или запутанные генетические регуляции). Какие бы стратегии ни стали доступными, наши результаты настойчиво подсказывают, что давление на их использование может быть велико. Суть нашей модели может захватывать нетипичных паразитов в социальных, технологических или экономических системах (например, когда компании получают прибыль от интеллектуальных усилий друг друга, когда хакеры используют вычислительную мощность без подозрений удаленных серверов [Barabasi et al., 2001[30]] или когда трейдеры предвидят ходы друг друга [Lewis, Baker, 2014[31]]). Паразиты также могли оказать влияние на эволюцию нейронных структур [del Giudice, 2019[32]], расширяя таким образом наши вопросы на машинное обучение и когнитивные процессы в целом. Недавний выдающийся успех Генеративно-состязательных сетей (Generative Adversarial Networks) [Goodfellow et al., 2014[33]] базируется на двух системах (с фиксированной сложностью), устанавливающих антагонистическую динамику, аналогичную нашей: одна сеть улучшает свою пригодность, обманывая другую искусственными данными, а другая становится более пригодной, научившись различать поддельные примеры. Наши результаты предоставляют возможность исследования появления все более сложных представлений в цифровых экосистемах, а также серьезные гипотезы о механизмах развития продвинутого когнитивного мышления в реальном мире.

## СПИСОК ЛИТЕРАТУРЫ